\documentclass[twocolumn,prd,noshowpacs,nofootinbib,amsmath,amssymb,superscriptaddress]{revtex4}
\usepackage{graphicx,epsfig,psfrag,bm,amssymb}
\usepackage{dcolumn}
\usepackage{bm}
\usepackage{mciteplus}
\usepackage{color}
\usepackage{mathrsfs,amsfonts,hepunits, color}
\usepackage{mciteplus}
\usepackage{tikz}
\usepackage{slashed}
\usetikzlibrary{arrows,shapes}
\usetikzlibrary{trees}
\usetikzlibrary{matrix,arrows} 				
\usetikzlibrary{positioning}				
\usetikzlibrary{calc,through}				
\usetikzlibrary{decorations.pathreplacing}  
\usepackage{pgffor}							

\usetikzlibrary{decorations.pathmorphing}	
\usetikzlibrary{decorations.markings}
\usetikzlibrary{snakes}

\tikzset{
    vector/.style={decorate, decoration={snake}, draw},
	provector/.style={decorate, decoration={snake,amplitude=2.5pt}, draw},
	antivector/.style={decorate, decoration={snake,amplitude=-2.5pt}, draw},
    fermion/.style={draw, postaction={decorate},
        decoration={markings,mark=at position .55 with {\arrow[draw]{>}}}},
    fermionbar/.style={draw, postaction={decorate},
        decoration={markings,mark=at position .55 with {\arrow[draw=black]{<}}}},
    fermionnoarrow/.style={draw},
    gluon/.style={decorate, draw,decoration={coil,amplitude=4pt, segment length=6pt}, line width=1},
    scalar/.style={dashed,draw, postaction={decorate},
        decoration={markings,mark=at position .55 with {\arrow[draw]{>}}}},
    scalarbar/.style={dashed,draw, postaction={decorate},
        decoration={markings,mark=at position .55 with {\arrow[draw]{<}}}},
    scalarnoarrow/.style={dash pattern = on 6 pt off 3 pt,draw},
    electron/.style={draw, postaction={decorate},
        decoration={markings,mark=at position .55 with {\arrow[draw]{>}}}},
	bigvector/.style={decorate, decoration={snake,amplitude=4pt}, draw},
	vectorscalar/.style={loosely dotted,draw, postaction={decorate}},
}
\RequirePackage{xspace}
\usepackage{relsize}
\usepackage{slashed}

\newcommand{\be}{\begin{eqnarray}}
\newcommand{\ee}{\end{eqnarray}}
\def\lsim{\mathrel{\rlap{\lower4pt\hbox{\hskip 0.5 pt$\sim$}}
    \raise1pt\hbox{$<$}}}                
\def\gsim{\mathrel{\rlap{\lower4pt\hbox{\hskip1pt$\sim$}}
    \raise1pt\hbox{$>$}}} 
    
\newcommand{\s}{{\rm s}}

\def\lsim{\mathrel{\rlap{\lower4pt\hbox{\hskip1pt$\sim$}}
    \raise1pt\hbox{$<$}}}
\def\gsim{\mathrel{\rlap{\lower4pt\hbox{\hskip1pt$\sim$}}
    \raise1pt\hbox{$>$}}}

\newcommand{\gev}{{\rm GeV}}

\begin{document}

\title{
The Galactic Center Excess from the Bottom Up}

\author{Eder Izaguirre}
\author{Gordan Krnjaic}
 \affiliation{Perimeter Institute for Theoretical Physics, Waterloo, Ontario, Canada    }
\author{Brian Shuve}
 \affiliation{Perimeter Institute for Theoretical Physics, Waterloo, Ontario, Canada    }
 \affiliation{Department of Physics and Astronomy, McMaster University,  
Hamilton, Ontario, Canada}

\begin{abstract}
It has recently been shown that dark-matter annihilation
 to bottom quarks provides a good fit to the galactic-center gamma-ray excess
 identified in the Fermi-LAT data. In the favored dark matter mass range $m\sim 30-40$ GeV, achieving
the best-fit annihilation rate $ \sigma v \sim 5\times 10^{-26} $ cm$^{3}$ s$^{-1}$ with perturbative couplings
requires a sub-TeV mediator particle that interacts 
 with both dark matter and bottom quarks. In this paper, we consider the 
 minimal viable scenarios in which a Standard Model singlet mediates $s$-channel interactions {\it only} between dark matter and
  bottom quarks, focusing on axial-vector, vector, and pseudoscalar couplings.
  Using simulations that include on-shell mediator production, we show that existing sbottom searches 
currently offer the strongest sensitivity over  a large region of the favored parameter space explaining the gamma-ray excess, particularly for axial-vector interactions. The 13 TeV LHC will be even more sensitive; however, it may not be sufficient to fully cover the favored parameter space, and the pseudoscalar scenario will remain unconstrained by these searches. 
We also find that direct-detection constraints, induced through loops of bottom quarks, complement collider bounds 
to disfavor the vector-current interaction when the mediator is heavier than twice the dark matter mass. 
We also present some simple models that generate pseudoscalar-mediated annihilation predominantly to bottom quarks. 
\end{abstract}

\maketitle

%
%

\section{Introduction}

Although dark matter (DM) constitutes roughly 85\% of the matter in our universe, its identity and 
interactions are currently unknown \cite{Beringer:1900zz}.
If DM annihilates to visible states, 
existing space-based telescopes may be sensitive to the flux of  
annihilation byproducts arising from regions of high DM density, including
  the galactic center (GC).

 Several groups have  confirmed a statistically-significant excess in the Fermi-LAT
 gamma-ray spectrum  \cite{Hooper:2010mq,Boyarsky:2010dr,Hooper:2011ti,Abazajian:2012pn, Hooper:2012sr,Gordon:2013vta,Abazajian:2014fta,Daylan:2014rsa, Huang:2013pda, Huang:2013apa}, originally identified in \cite{Goodenough:2009gk}. 
The excess is largely confined to an angular size of $\lsim 10^\circ$ with respect to the GC, exhibits spherical symmetry,
and is uncorrelated with the galactic disk or Fermi bubbles \cite{Daylan:2014rsa}.
 While this excess may still be astrophysical in origin, potentially due to an unusual population of millisecond pulsars \cite{Gordon:2013vta},
its energy spectrum and spatial distribution are well modeled by an Navarro-Frenk-White profile \cite{Navarro:1996gj} of dark matter particles $\chi \bar \chi$ annihilating to $b \bar b$ with 
mass and cross section \cite{Abazajian:2014fta} 
\be
\label{eq:rates}
\langle \sigma v \rangle &=& (5.1 \pm 2.4)  \times 10^{-26}~ \cm^3 \s^{-1}~~, \\ 
 m_\chi &=& 39.4~(^{+3.7}_{-2.9} {~ \rm stat.}) (\pm 7.9 {\rm ~sys.})~ \gev~~,  
\ee
which are compatible with a DM abundance from thermal freeze-out. 

Recent work has presented the collider and direct-detection constraints on this interpretation assuming  flavor-universal 
and  mass-proportional couplings to SM fermions \cite{Alves:2014yha, Berlin:2014tja}; these analyses apply collider bounds on DM production assuming a contact interaction between dark and visible matter. 
The analyses in \cite{DiFranzo:2013vra,Berlin:2014tja,AgrawalBatellLinHooper} also study simplified models of DM annihilation mediated by color-charged $t$-channel mediators.  For perturbative interactions, Eq.~(\ref{eq:rates}) implies 
 that the mediator mass is below a TeV, so it  can be produced on-shell at the Large Hadron Collider (LHC) 
 and decay to  distinctive final states with a mixture of $b$-jets and missing energy ($\displaystyle{\not}{E}_T$).
At direct-detection experiments, this mediator can also be integrated out to induce dark matter scattering
 through loops of $b$ quarks that exchange photons or gluons with nuclei. 
 Up to differences in Lorentz structure, these processes are generic predictions of any model that explains the Fermi anomaly;
 however for light mediators ($<2m_\chi$), it is possible to evade collider
 searches \cite{Boehm:2014hva}.

 In this paper, we study the scenario with an $s$-channel mediator that predominantly couples to 
 $b$ quarks and focus on the regime in which the mediator is $\gsim 100$ GeV and can decay to pairs of DM particles.
  The mediator can be produced in processes involving $b$ quarks, and its decays 
 yield final states with $b$ jets and/or missing energy.  We extract constraints from LHC searches for new physics in the 
 $b \bar b + {\displaystyle{\not}{E}_T}$ final state and explore the sensitivity of a proposed mono-$b+\displaystyle{\not}{E}_T$ analysis \cite{Lin:2013sca}. 
We find that large regions of favored parameter space are
 excluded by existing 8 TeV sbottom searches, whose sensitivity is projected to improve at 13 TeV. The mono-$b$ analysis is expected to be comparable at 8 TeV and set stronger constraints at 13 TeV. We also clarify the LUX limits \cite{Akerib:2013tjd} on scattering through loops of $b$ quarks and find strong bounds on the parameter space of vector-mediators that explain the Fermi excess.  

The organization of the paper is as follows: in section \ref{sec:general} we discuss a set of possible minimal interactions that can explain the GC excess. In section \ref{sec:ddres}, we consider direct-detection, resonance, and Higgs search constraints on these scenarios. In section \ref{sec:LHC}, we show the constraints on these DM interpretations from sbottom LHC searches, which allow for a possible independent, complementary confirmation of the GC excess. We also estimate the reach of a mono-$b$ search at 8, and extend our results for both searches to 13 TeV. In section \ref{sec:concrete}, we outline concrete 
models that generate a pseudoscalar mediated annihilation, which is the least constrained of all possible operators that can explain the gamma-ray anomaly.

%
%
 
\section{Annihilation Operators} \label{sec:general}


In the simplest models, dark matter can consist of fermions, scalars, or vector bosons. 
To narrow the scope of our investigation without essential loss of generality, 
we consider only parity-conserving interactions between dark and visible matter. For scalar DM, the leading-order interaction
with $b \bar b$ is either ruled out by direct detection bounds or the annihilation is 
$p$-wave suppressed \cite{Berlin:2014tja}, so achieving the rate in Eq.~(\ref{eq:rates}) in the latter case requires non-perturbative couplings. For vectors, protecting the DM from prompt decays requires nontrivial model building, so for simplicity we omit this possibility.
Thus, for the remainder of this paper we focus exclusively on Dirac fermion DM candidates; Majorana particles are qualitatively 
similar and the collider bounds are expected to be comparable.  

We separately consider the following interactions  \vspace{-0.05cm}
 \be 
  {\cal L}_{U} &=& \left(g_\chi \bar \chi \gamma^\mu \gamma^5 \chi +  g_b  \bar b \gamma^\mu \gamma^5  b \right)U_\mu  ~~,~~ \label{eq:axialvector} \\ 
  {\cal L}_V &=& \left(g_\chi \bar \chi \gamma^\mu \chi +  g_b  \bar b \gamma^\mu  b \right)V_\mu  ~~,~~  \label{eq:vector}\\ 
 {\cal L}_a &=& i\left(g_\chi \bar \chi \gamma^5\chi +  g_b  \bar b \gamma^5  b \right)a  ~~,~~ \label{eq:pseudoscalar}
 \ee
where $U,V$ and $a$ are axial-vector, vector, and  pseudoscalar fields that mediate $s$-channel $\chi$ and $b$ interactions. We assume the mediator is a singlet under SM  gauge interactions, and thus we do not address $t$-channel mediators in this article (see Ref.~\cite{Berlin:2014tja} for constraints on the latter). Our collider and direct-detection constraints assume only these interactions between the SM and DM. To leading order in velocity, the annihilation cross sections are 
\be
\label{eq:annihilation}
\langle  \sigma v \rangle_{U} &\simeq& \frac{N_c}{2\pi} \frac{  (g_\chi g_b)^2  m_b^2 (1-4m_\chi^2/m_U^2)^2  \sqrt{1 - m_b^2/m_\chi^2 }  }{ \left( m_{U}^2   - 4 m_\chi^2 \right)^2 + m_{U}^2 \Gamma_U^2} 
  ,~~~~
\\
\langle  \sigma v \rangle_V &\simeq& \frac{N_c}{\pi} \frac{ (g_\chi g_b)^2 \, m_\chi^2 (1+m_b^2/2 m_\chi^2) \sqrt{1 - m_b^2/m_\chi^2 }} {  \left( m_V^2   - 4 m_\chi^2 \right)^2 + m_V^2 \Gamma_V^2  }    \label{eq:annihilation2}
  ~, ~~\\   
  \langle \sigma v\rangle_{a}  &\simeq & \frac{N_c}{2\pi} \frac{ (g_\chi g_b)^2 \, m_\chi^2\sqrt{1 - m_b^2/m_\chi^2} } {  \left( m_a^2   - 4 m_\chi^2 \right)^2 + m_a^2 \Gamma_a^2  }  ~~,  \label{eq:annihilation3}
\ee
where $N_c = 3$ is the number of colors. For $g_b g_\chi = 1$, the best fit values from Eq.~(\ref{eq:rates}) imply
mediator masses in the few-hundred GeV range,  
which are light enough to be accessible with a combination of experimental strategies.  For lighter mediators, the constraints due to direct-detection and collider experiments are quite weak and consistent with a DM interpretation of the gamma-ray excess \cite{Boehm:2014hva}. In sections \ref{sec:ddres} and \ref{sec:LHC}, we discuss the various constraints on the three sets of interactions (Eqs.~\ref{eq:axialvector},\ref{eq:vector},\ref{eq:pseudoscalar}) from direct-detection experiments and collider searches.


\section{Direct Detection and Resonance Searches}
\label{sec:ddres}

\subsection{Direct Detection}
\begin{figure}[t!]
 \begin{minipage}[h]{1.\linewidth}
	\includegraphics[width=7.cm]{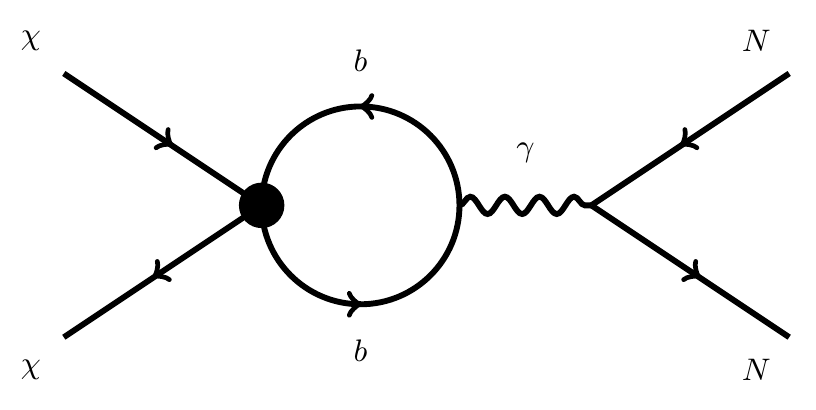}
     \end{minipage}
  \caption{
  Minimal loop-induced nucleon scattering at direct-detection experiments for an $s$-channel mediator particle between $\chi$ and $b$. The mediator has been integrated out.}
\label{fig:loopscat}
\end{figure}

The LUX experiment currently places the strongest limit 
on spin-independent $\chi$-nucleon interactions over the $m_\chi \sim 10 -100$ GeV range,  at $\sigma_{SI} \lsim 10^{-46}$ cm$^2$ \cite{Akerib:2013tjd}. Although we assume that none of $a$, $U$, or $V$ couple directly to light quarks, 
it is still possible for DM to induce elastic nuclear scattering through a loop of $b$ quarks
 as depicted in Fig.~\ref{fig:loopscat}. 
 The cross section for the vector-mediated process in the leading-log 
 approximation is \cite{Kopp:2009et}
 \be
 \frac{d\sigma}{dE}  = \frac{(g_b g_\chi)^2 \,m_{\rm T}}{18 \pi v^2  \, m_V^4  } \left(\frac{\alpha  Z}{\pi}\right)^2 F^2(E) \left[ \log\left(\frac{m_b^2}{m_V^2} \right) \right]^2,
 \ee
 where  $E$ is the nuclear recoil energy, $m_{\rm T}$ is the mass of a target nucleus, $Z$ is the target's electric charge,
   $v$ is the relative velocity, and $F$ is the Helm form factor \cite{Helm:1956zz}.
  The scattering rate in units of counts/day/keV/kg detector-mass is 
\be
\frac{dR}{dE} = \frac{\rho_{\chi}}{m_\chi m_{\rm T}} \int_{v_{\small \rm min}(E)}^{v_{\small \rm esc}} d^3 v  f_{\odot}(\vec v, v_0) \, v \,\frac{d\sigma}{dE}  ~,~
\ee
where $\rho_\chi = 0.3$ GeV/cm$^3$ is the local DM mass density,  $v_{\small \rm min}(E) = \sqrt{ m_T E/2\mu}$ 
is the minimum DM velocity required to induce a nuclear recoil of energy $E$,  $\mu = m_\chi m_{\rm T} /(m_\chi + m_{\rm T})$ is the reduced mass,  
$v_{\small  \rm esc} \approx 550 $ km/s is the halo escape velocity, and $v_0 = $ 220 km/s is the mean 
local DM velocity.  Here, $f_{\odot}(\vec v,v_0)$ is the local DM velocity distribution in the detector frame, 
which is obtained from a Maxwellian distribution in the galactic rest frame
boosted by the Earth's velocity with respect to the halo.   

Using the LUX limits and detection efficiencies \cite{Akerib:2013tjd}, we find the  $(\bar \chi \gamma^\mu \chi)(\bar b \gamma_\mu b)$ interaction
is disfavored over much of the $m_V > 2m_\chi$ range
 as shown in Figure \ref{fig:luxbound}. 
  For the pseudoscalar and axial-vector interactions in Eq.~(\ref{eq:axialvector}) and Eq.~(\ref{eq:pseudoscalar}), LUX places no relevant constraint
 since the one loop diagram depicted in Fig.~\ref{fig:loopscat} vanishes and the leading process is spin-dependent\footnote{
 Our LUX limit on the $(\bar \chi \gamma^\mu \chi)(\bar b \gamma_\mu b)$ interaction (green curve in Figure \ref{fig:luxbound}, color online)
 disagrees with the bounds on spin-1 $s$-channel interactions in Figure 3 of \cite{Berlin:2014tja}, which cites \cite{Agrawal:2011ze} 
 for the loop induced scattering cross section. However, the diagrams 
  calculated in Appendix A of \cite{Agrawal:2011ze} feature a $t$-channel $\chi b$ interaction, whereas the 
  vector-vector interaction with an $s$-channel mediator arises from the 
  process depicted in Figure \ref{fig:loopscat}, which sets a stronger bound on this process.
}.


\subsection{Resonance Searches}
\label{sec:resonances}

We consider constraints on non-standard $b$-jet production in the context of dijet resonances and non-SM Higgs searches. 
Mediator production at hadron colliders proceeds  via 
 \be \vspace{-0.1cm}
 pp &\rightarrow& U/V/a \rightarrow b \bar{b}\label{eq: dijet searches},
 \ee
 and  yields dijet resonances. 
The best limits are from UA2 and Tevatron dijet searches \cite{Dobrescu:2013cmh}, which bound a 
universal $Z^\prime$ coupling by $\lsim 0.5$ over the $m_{Z^\prime} \in$ 100-1000 GeV range.  In our scenarios of interest, the 
mediators couple only to $b$ quarks, so the production rate is suppressed  by parton distribution functions
and there is no constraint for perturbative mediator couplings to $b$. 

Similarly, the CMS search for non-standard Higgs sectors is sensitive to 
final states with 3 or more jets  \cite{Chatrchyan:2013qga}, which can arise in our scenarios of 
interest via
\be
pp &\rightarrow& (U/V/a\rightarrow b \bar{b}) \label{eq: higgs searches} + b ~{\rm jets}.
\ee
 Simulating inclusive $U,V,$ and $a$,  
production using {\tt Madgraph 5}  \cite{Alwall:2011uj}, and applying the CMS limits from \cite{Chatrchyan:2013qga}, we find this bound to be comparable to the sbottom and mono-$b$ searches considered in Section \ref{sec:LHC} for pseudoscalars when $g_b=g_\chi$ (see Fig.~\ref{fig:moneyplot}). A similar bound is expected for axial-vector and vector mediators; however, the different kinematics of the (axial-)vector final states prevent a direct application of the bound, and the Higgs searches are anyway subdominant to the sbottom constraints for these mediators. In the $|g_\chi| \ll |g_b|$  limit, the multi-$b$ Higgs search no longer applies, as the coupling to $b$ quarks is relatively suppressed.

%
%

\section{ Collider DM Searches }
\label{sec:LHC}
Collider studies of DM production in association with SM particles have proliferated vastly in recent years 
\cite{Petriello:2008pu,Gershtein:2008bf,Cao:2009uw,Beltran:2010ww,Goodman:2010yf, Bai:2010hh,Goodman:2010ku,Fox:2011fx,Fortin:2011hv,Frandsen:2011cg,Rajaraman:2011wf,Fox:2011pm,Goodman:2011jq,Shoemaker:2011vi,An:2012va,Fox:2012ee,Frandsen:2012rk, Bai:2012xg,Cotta:2012nj,Carpenter:2012rg,Dreiner:2013vla,Busoni:2013lha,Yu:2013aca,Profumo:2013hqa,An:2013xka,
AgrawalBatellLinHooper}.
 In this section, we consider interactions in which DM couples predominantly to $b$ quarks through the axial-vector, vector, and  pseudoscalar interactions (Eqs.~\ref{eq:axialvector}-\ref{eq:pseudoscalar}). Although DM annihilation in the GC is well approximated by the contact-interaction limit for $m_\chi \ll m_{a, V, U}$, 
 the preferred mediator-masses are of order a few-hundred GeV for perturbative couplings. Therefore, due to the high partonic center-of-mass energies at the LHC, the effective theory description \cite{Busoni:2013lha,Buchmueller:2013dya,Berlin:2014tja} is not applicable. In this section, we show the LHC's sensitivity to on shell production of pseudoscalar, vector, and axial-vector mediators, highlighting the parameter space suggested by the Fermi excess.  The generic DM production process at  the LHC is 
\begin{eqnarray}
pp &\rightarrow& (U/V/a\rightarrow \chi \bar{\chi}) \label{eq: sbottom searches} + X_{\text{sm}},
\end{eqnarray}
where $X_{\text{sm}}$ can be any multiplicity of SM final states and the $U/V/a \to \chi \bar \chi$ decay yields missing energy in the final state. There are several scenarios to consider, depending on the nature of the additional SM final states, $X_{\text{sm}}$, produced in association with $U/V/a$. For $X_{\text{sm}}=W^{\pm}$, $X_{\text{sm}}=Z^0$, or $X_{\text{sm}} = j (\neq b)$, the signal could appear in the mono-$X_{\text{sm}}+\displaystyle{\not}{E}_T$ searches \cite{Alves:2014yha}. 

\begin{figure}[t!]
 \begin{minipage}[h]{1.\linewidth}
	\includegraphics[width=7.7cm]{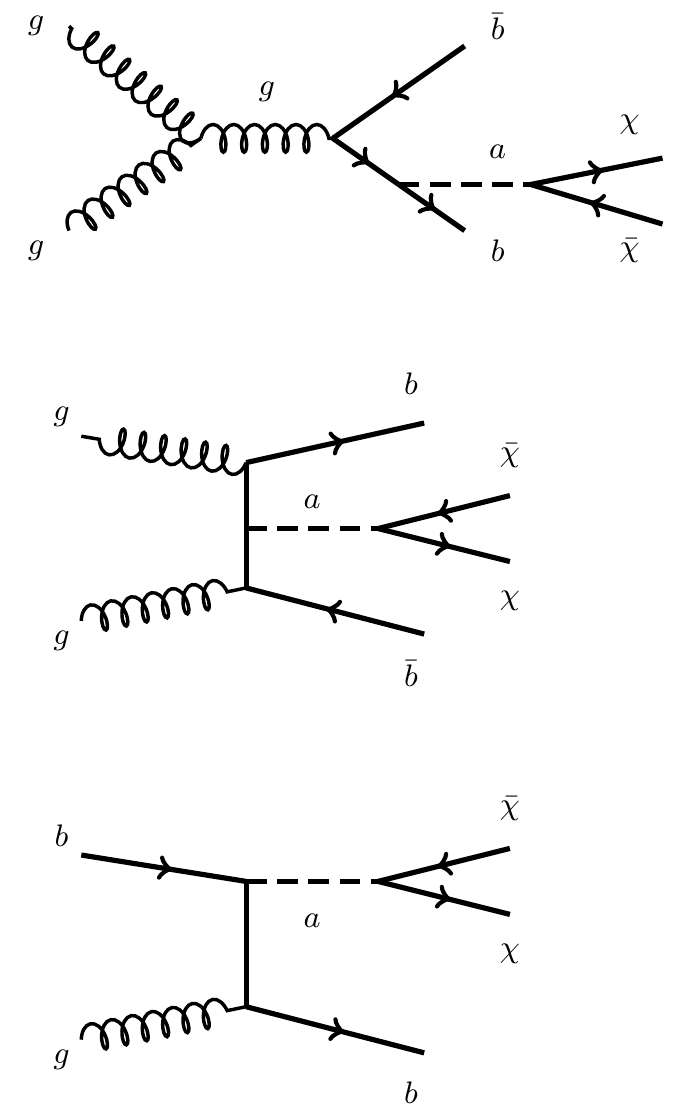}
     \end{minipage}
  \caption{Example diagrams for $p p \to b \bar b~ \chi \bar \chi$. If kinematically allowed, the dominant process is $p p \to b \bar b \, (a \to \chi \bar \chi)$ 
  which suffers less phase space suppression.
}
\label{fig:diag}
\end{figure}

However, the best sensitivity to these signals utilizes the power of $b$-tagging, since mediator production is almost always accompanied by at least one associated $b$-quark. Fig.~\ref{fig:diag} depicts representative Feynman diagrams that give rise to  $b$ quarks and missing energy from DM production processes.
 Ref.~\cite{Lin:2013sca} proposed a mono-$b$ analysis which can set strong constraints on the topologies considered in this article by looking for a $b$-tagged jet and significant missing energy. To date, this analysis has not yet been performed.

A central result of our paper is that strong bounds can already be set with existing LHC  sbottom searches in the $2b+\displaystyle{\not}{E}_T$ channel. 
We note that this final state was considered by \cite{Berlin:2014tja} in the context of the pair-production of a colored $t$-channel mediator between $b$ quarks and DM. 
Here, we show that the sbottom searches also place constraints on {\it $s$-channel} mediators that are uncharged under the SM and are produced only
 through the interaction responsible for $\chi \bar \chi \to  U/V/a \to b \bar b $ annihilation. 

Our Monte Carlo calculations of the SM backgrounds for the mono-$b+\displaystyle{\not}{E}_T$ and $2b+\displaystyle{\not}{E}_T$ final states were done in \texttt{Madgraph 5} \cite{Alwall:2011uj}. We include samples of the dominant SM processes, namely $V+\text{jets}$, and $t\bar t + \text{jets}$, which are matched with the $k_\perp$-shower scheme \cite{Alwall:2008qv}. Next-to-leading-order (NLO) k-factors for the backgrounds are calculated with \texttt{MCFM} \cite{Campbell:2002tg, Campbell:2012uf}. The pseudoscalar and axial vector operators are also simulated in \texttt{Madgraph 5} with a user-defined model. Showering and additional jets from initial- and final-state radiation  are generated in \texttt{Pythia 6.4} \cite{Sjostrand:2006za}, with a detector simulation done in \texttt{PGS 4} \cite{PGS}. The \texttt{PGS} version used in this study is modified from the standard version \cite{Essig:2011qg}; in this modified \texttt{PGS}, the truth $b$ and $c$ tagging was improved, and the anti-$k_{\rm T}$ clustering was incorporated from \cite{Cacciari:2008gp}. This study uses an $R=0.4$ clustering radius. We validated the backgrounds simulated in this study by reproducing the expected background yield in the signal regions of the ATLAS sbottom search \cite{Aad:2013ija}  to within 20\%-30\%.

\begin{figure}[t]
 \begin{minipage}[h]{1.\linewidth}
	\includegraphics[width=8.7cm]{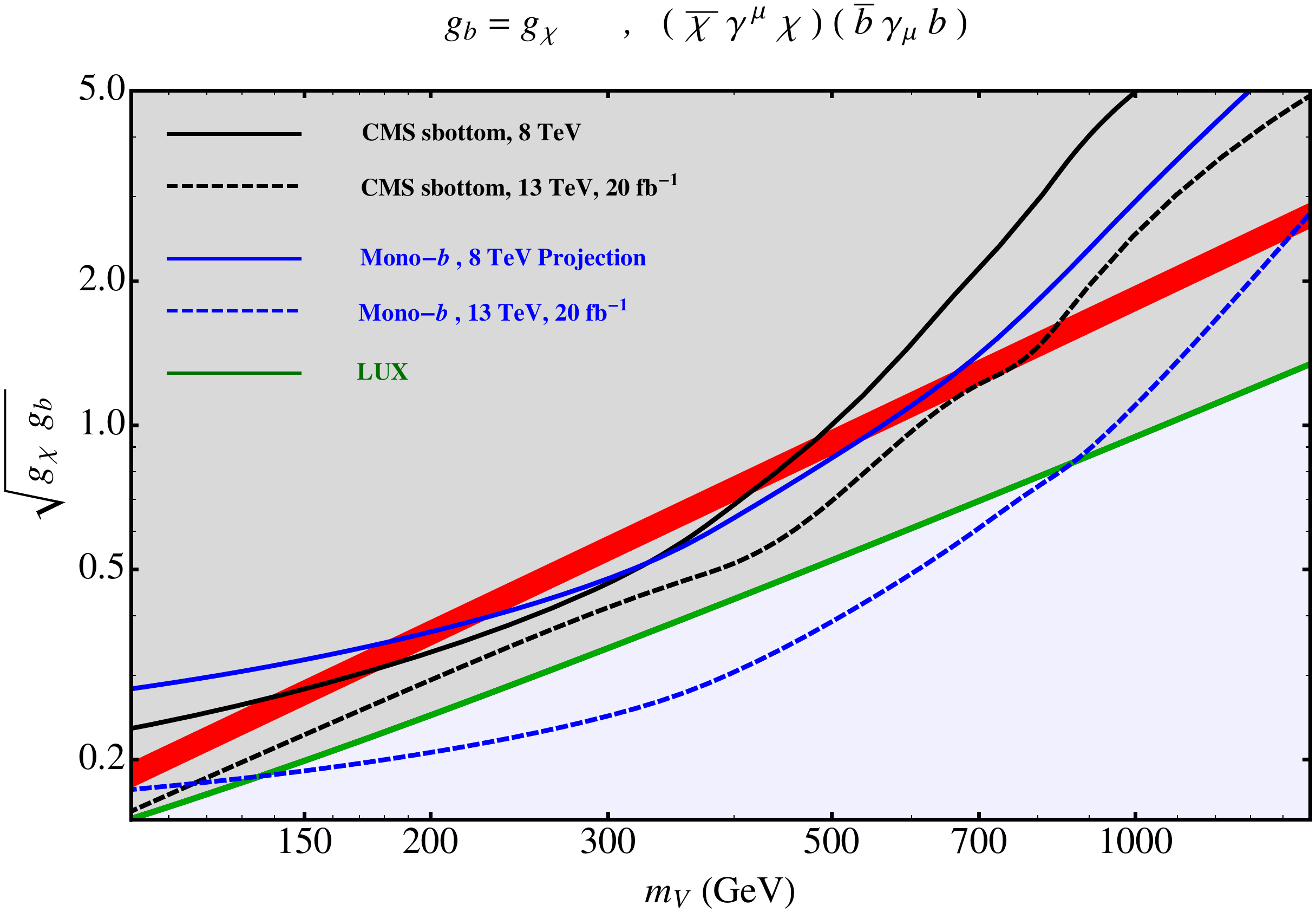}
     \end{minipage}
     
    \vspace{0.1cm} 
     
 \begin{minipage}[h]{1.\linewidth}
	\includegraphics[width=8.7cm]{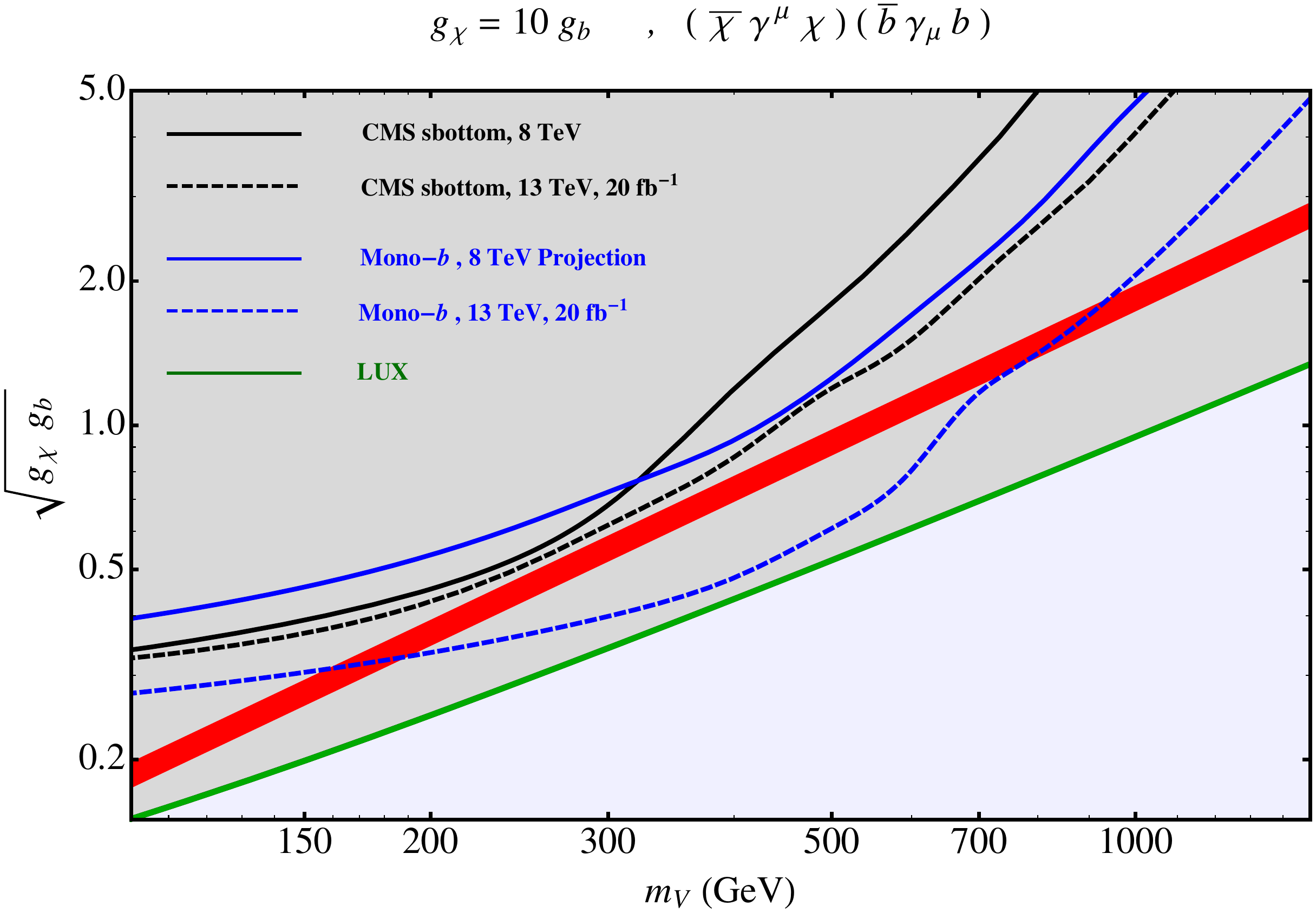}
     \end{minipage}
     
  \caption{
  Direct-detection and collider constraints on the vector-mediated scenario.
The red band (color online) is the favored region for  for the $\chi \bar \chi \to V^* \to b \bar b$ annihilation 
in the GC \cite{Abazajian:2014fta}.     The gray excluded region is extracted from the 8 TeV CMS sbottom search \cite{cms8} -- comparable limits arise from the 
       ATLAS sbottom search in \cite{Aad:2013ija} -- and the dashed blue line shows the projected sensitivity of the mono-$b$ 
       search using the cuts proposed in \cite{Lin:2013sca} at $\sqrt{s} =  8$ TeV.   
The green curve is the LUX bound using limits and efficiencies from \cite{Akerib:2013tjd}.  
}
\label{fig:luxbound}
\end{figure}

\begin{figure}[t!]
        \centering	\includegraphics[width=8.5cm]{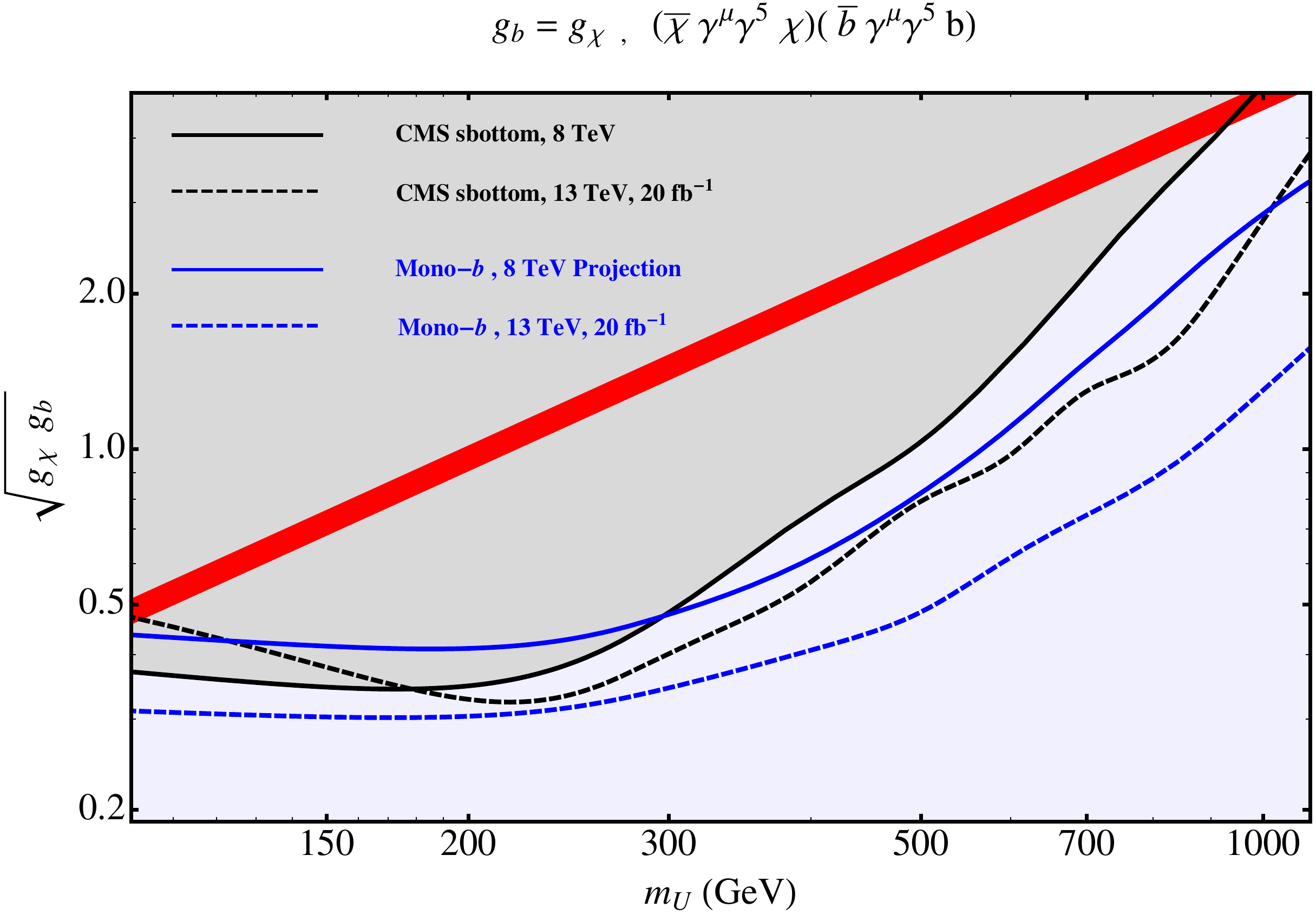}
        
        \vspace{0.2cm}
        
        \centering	\includegraphics[width=8.5cm]{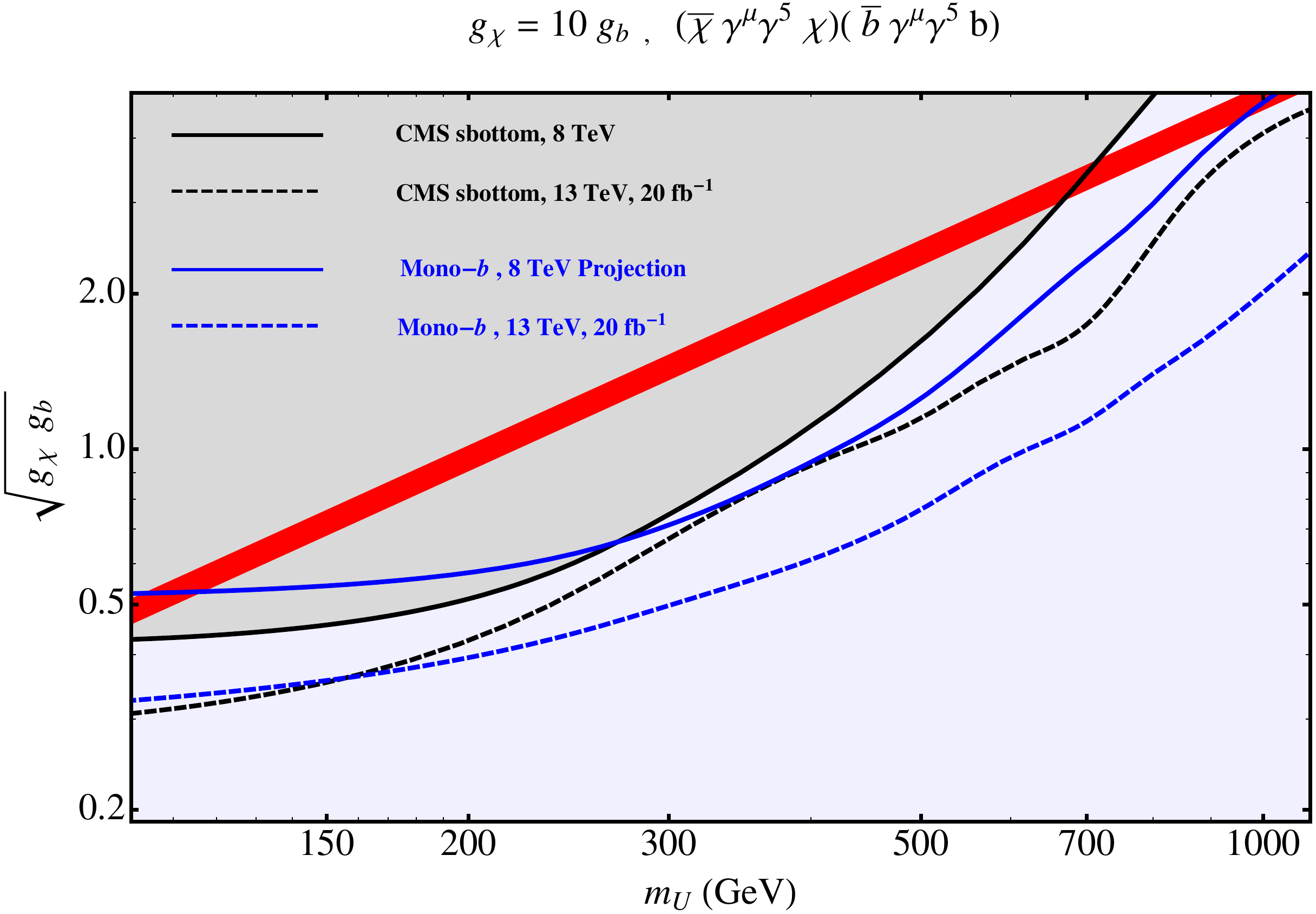}

       \caption{
       Parameter space for the axial-vector-mediated scenario with constraints from the same searches and simulation details described in Fig.~\ref{fig:luxbound}.  For
       this interaction, the LUX bound for scattering through a $b$-quark loop is not significant as the leading scattering process is spin-dependent. 
       }
\label{fig:axialvector}
\end{figure}

\begin{figure}[t!]
        \centering	\includegraphics[width=8.5cm]{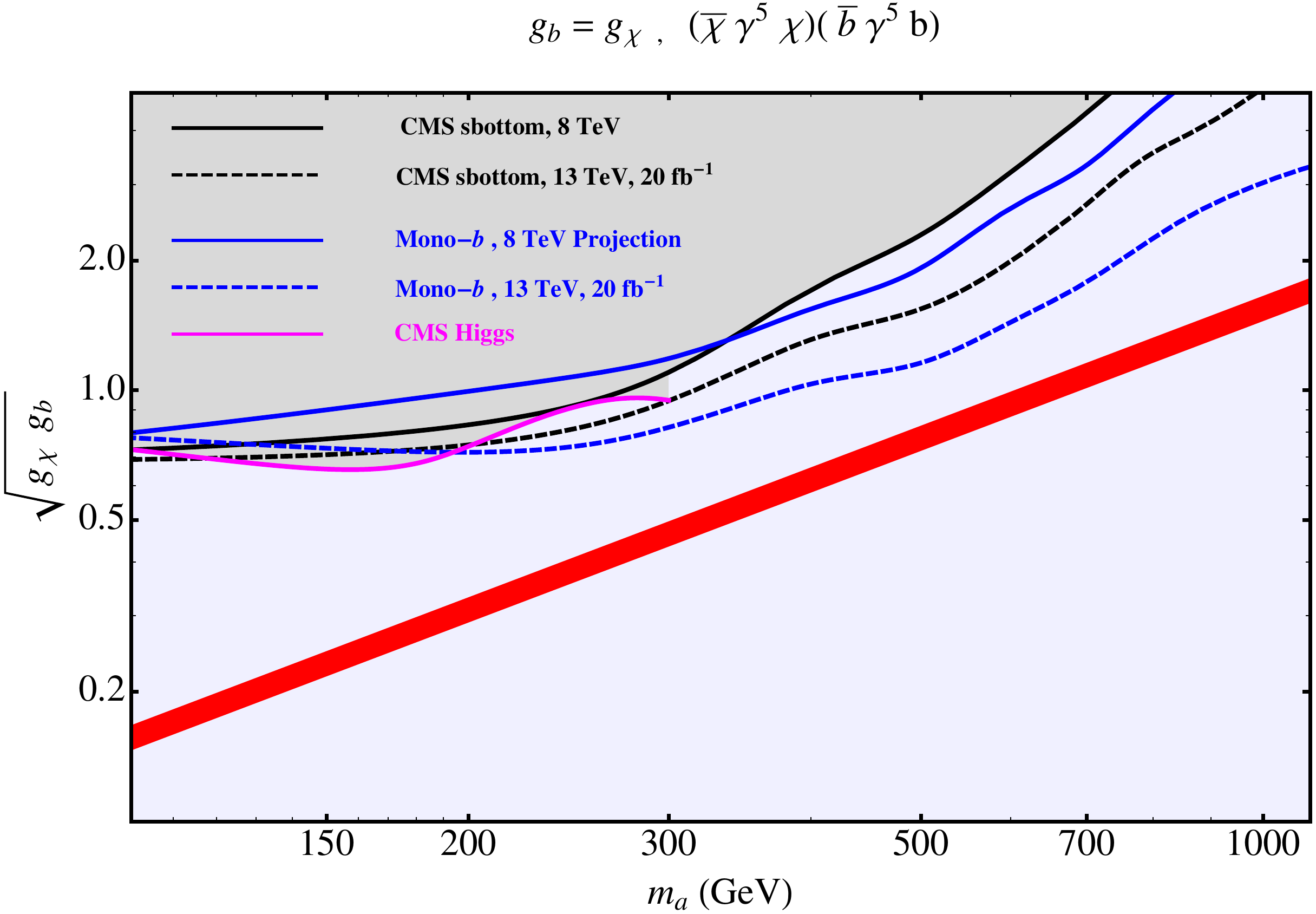}
        
                \vspace{0.2cm}

        \centering	\includegraphics[width=8.5cm]{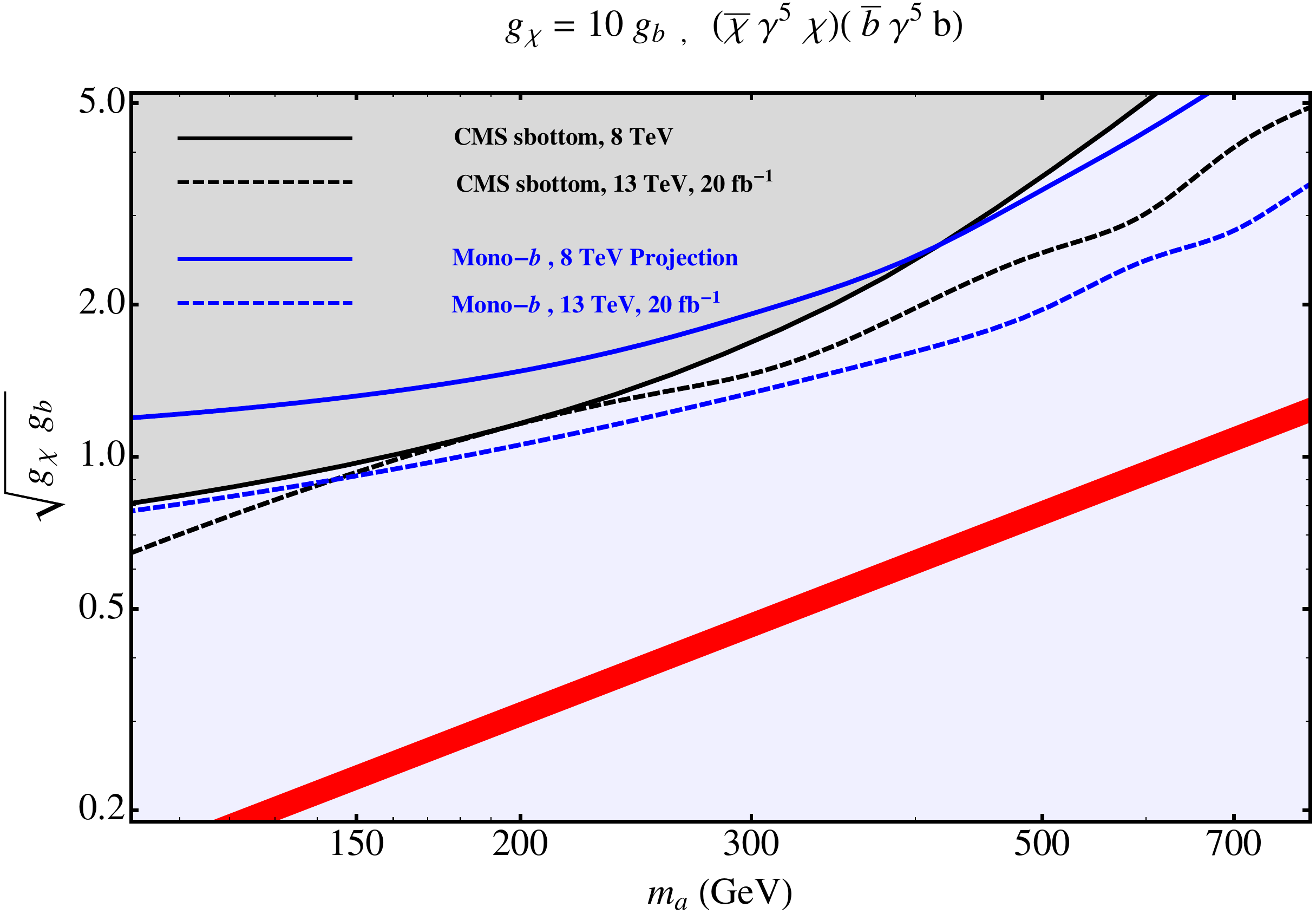}

       \caption{
       Parameter space for the pseudoscalar-mediated scenario with constraints from the same searches and simulation details described in Fig.~\ref{fig:luxbound}. As with the 
       axial-vector mediator, scattering at direct-detection experiments through a $b$ loop is not constraining as the leading interaction is spin-dependent. Here
       we also include a constraint from the CMS Higgs search from \cite{Chatrchyan:2013qga}}
\label{fig:moneyplot}
\end{figure}

Our main results are encapsulated in Figures \ref{fig:luxbound}, \ref{fig:axialvector}, and \ref{fig:moneyplot}, which 
present the constraints on the vector, axial-vector, and pseudoscalar operators, respectively. All three Figures show constraints from collider production and direct detection
for $g_\chi = g_b$ and $g_\chi = 10 g_b$; the gray region in each plot is ruled out by existing searches, while the other curves 
show projections of potential future sensitivities. 
For clarity of presentation, we emphasize the CMS  sbottom search \cite{cms8}, which already constrains a large region of parameter space 
for several scenarios, though comparable sensitivity is achieved with the corresponding ATLAS analysis \cite{Aad:2013ija}. The LHC is expected to have already put strong constraints on vector and axial-vector interactions for a range of parameter space that can explain the Fermi gamma-ray excess. In the pseudoscalar scenario, however, the LHC constraints on the UV completion of this operator are not expected to robustly test the gamma-ray excess preferred parameter space. 

 In Figures \ref{fig:luxbound}, \ref{fig:axialvector}, and \ref{fig:moneyplot}, we also show our estimated sensitivity for 95\% confidence level (CL) exclusion from future sbottom searches at 13 TeV, assuming the same selection criteria from the analysis at 8 TeV, with the addition of an optimization over missing transverse energy, $\displaystyle{\not}{E}_T$. The expected bounds that we draw at 13 TeV assume a systematic uncertainty of 10\%. At 20 fb$^{-1}$, the signal regions we consider are already systematics-dominated, and longer running will not necessarily improve the bounds.

We also show the sensitivity for 95\% CL exclusion from a mono-$b+\displaystyle{\not}{E}_T$ analysis  proposed in \cite{Lin:2013sca} using the $b$-tagging working point from the CMS sbottom search \cite{cms8}. This analysis offers more optimal coverage at high mediator masses, where the signal benefits from a hard radiated jet whose recoil boosts the $\chi\bar\chi$ system and consequently enhances the missing energy spectrum \cite{Alwall:2008ve}. 

In Figs.~\ref{fig:luxbound}, \ref{fig:axialvector} and ~\ref{fig:moneyplot}, we show how the bounds compare for  $g_\chi = g_b$ and $g_\chi = 10 g_b$. As we show in Section~\ref{sec:concrete}, the $g_b$ coupling is typically smaller than the $g_\chi$ coupling. For $g_\chi \gg g_b$, the bounds from LHC searches are weakened, as the rate for radiating off an on-shell mediator gets smaller. 

In summary, we find that LHC searches with $b$ jets and missing energy are excellent probes of interactions responsible for the GC excess, particularly for interactions mediated by axial-vectors and vectors. In such scenarios, most of the parameter space with $m_{U,V}>2m_\chi$ is already excluded or will be in early 13 TeV running. However, pseudoscalars currently evade all such constraints and will be challenging to probe at 13 TeV with heavy flavor + DM searches.

%
%

\section{Beyond the Minimal Interaction} \label{sec:concrete}

\subsection{Pseudoscalar Mediated Models}
In this section, we study concrete models that give rise to $\bar \chi\chi \to \bar b b$ annihilation 
with pseudoscalar mediators. Our emphasis is motivated both by the larger allowed parameter space 
that remains for this scenario, and the difficulty of constructing viable vector and axial-vector interactions 
that give rise to appreciable annihilation rates. Note that the interactions in Eq.~(\ref{eq:pseudoscalar}) are not permitted prior to electroweak symmetry breaking  as the left- and right-handed bottom quarks have different gauge charges. Therefore, we generically find that  $g_\chi \gg g_b$ for a singlet mediator,
 as the coupling to the visible sector is often accompanied by some source of suppression (mixing angles, higher-dimensional operators, etc.).


\subsubsection{Two-Higgs Doublet Model with a Singlet}
\label{sec:2HDM}

A pseudoscalar with the interactions in Eq.~(\ref{eq:pseudoscalar}) can arise in a two-Higgs doublet model (2HDM)
with an additional complex-scalar singlet. Two-Higgs doublet models typically induce flavor-changing neutral currents (FCNCs) that are
strongly constrained unless each set of fermions couple predominantly to only one of the Higgs doublets \cite{Glashow:1976nt}.
We consider a scenario analogous to \cite{Batell:2012mj}, where one of the Higgs doublets, which we refer to as $H_u$, couples to the up-type quarks and the leptons, while the
other doublet, which we refer to as $H_d$, couples to the down-type quarks.
The Lagrangian for this scenario contains 
\be
{\cal L} \supset  g_\chi {\cal S} \bar \chi  \chi + {\lambda^{ij}_d} \bar Q^i H_d d_R^j-  \mu\, {\cal S} H_u H_d + h.c.~,
\ee
where $\chi$ is Dirac fermion uncharged under the SM,  ${\cal S} = (\phi + i a)/\sqrt{2}$ is a complex scalar and $H_{u,d} =  (h_{u,d} +ia_{u,d})/\sqrt{2}$. In the $\tan\beta \equiv v_u/v_d \gg1 $ limit, $v_u \approx v = 246$ GeV, the down-type Yukawa coupling $\lambda_d$ is of order one,  and $\cal S$ mixes predominantly with the down-type Higgs.  

Assuming $CP$-conservation, the scalars and pseudoscalars mix 
separately and acquire identical off-diagonal mass terms $ \sim \mu\,v/\sqrt{2}$. This mixing induces
both the desired $\chi \bar \chi \to a\to b \bar b$ annihilation, as well as scalar-mediated spin-independent
 scattering at direct-detection experiments, which is strongly constrained. Both processes are proportional to the mixing angles, which scale approximately as
 \be
 \sin \theta_a \sim \frac{ \mu v }{  \sqrt{2}\mu v+m^2_{a} + m^2_{a_{d}}   } ~,~\\ 
 \sin \theta_\phi \sim \frac{\mu v}{  \sqrt{2}\mu v +  m^2_{\phi} + m^2_{h_{d}}  } ~,~
 \ee 
 in the limit where one mass term dominates each of the numerator and denominator. $m_{a,\phi,a_d,h_d}$ are the tree-level mass terms prior to electroweak symmetry breaking. 

In the absence of tuning, the lightest scalar and pseudoscalar have comparable tree-level masses and there is 
generic tension between ensuring a $\lsim$ TeV pseudoscalar with a large mixing angle to explain the Fermi excess, 
and keeping at least one scalar component above $\gsim$ TeV to suppress elastic spin-independent scattering at LUX \cite{Akerib:2013tjd}. To alleviate this tension, we can make the mixing angles hierarchical by ensuring $m_{a}\sim m_{a_d}$ and 
 $m_{\phi} \gg m_{\phi_d}$, which implies a tuning in the $\cal S$ masses. The most general $CP$-conserving mass terms are 
\be
-{\cal L}_m \supset \mu_1^2 |{\cal S}|^2 + \mu_2^2 \,\mathrm{Re}({\cal S}^2), 
\ee 
which yield tree-level scalar and pseudoscalar mass-terms 
\be
m_a^2 &=& \mu_1^2 - \mu_2^2,\\
m_\phi^2 &=& \mu_1^2 + \mu_2^2.
\ee
These masses can be split, given a degeneracy of $\mu_1$ and $\mu_2$. To quantify the necessary hierarchy,
 let $m_a = x m_\phi$, where 
\be
x \equiv \sqrt{\frac{(\mu_1-\mu_2)(\mu_1+\mu_2)}{(\mu_1^2+\mu_2^2)}}.
\ee
Thus,  the splitting has to be tuned by a factor $x^2$. For large pseudoscalar mixings $\mu v\sim m_a\sim$ few hundred GeV, the mass ratio
$m_a/m_\phi\gtrsim10$ is required to evade LUX bounds on $\phi$-mediated scattering. This corresponds to a tuning of order $ x^2 \sim$ 1\%.


\subsubsection{Vectorlike Quarks}
It is also possible to induce the pseudoscalar couplings in Eq~(\ref{eq:pseudoscalar}) without extending the Higgs sector.
Consider the SM with an additional singlet pseudoscalar, $a$, and three generations of vectorlike quarks, $\Psi_i$, with charge $(3,2)_{\frac{1}{6}}$ under 
$SU(3)_c \times SU(2)_L \times U(1)_Y$. Up to field redefinitions, the most general renormalizable interactions are
\be
\mathcal L\supset y_{1ki} a \bar\Psi_{k} \gamma^5 Q_i + y_{2kj} \bar \Psi_k H d^j_{R}   + M_{\Psi,k} \bar\Psi_k\Psi_k~~,
\ee
where $H$ is the SM Higgs doublet. Integrating out $\Psi$ yields the effective interaction
\be
\mathcal L_{eff} = \frac{y_{ij}}{M_\Psi} \bar{Q}^i\gamma^5 d^j_{R} H a  \to \frac{y_{ij} v}{\sqrt{2} M_\Psi} \bar{Q}^i\gamma^5 d^j_{R} a ~,
\ee
where we define  $y_{ij} \equiv 
\sum_k y_{1ki}  y^*_{2kj}$. The effective Yukawa coupling must be  aligned with the down-type Yukawa matrix to avoid FCNCs. 

Since $v/\sqrt{2}\approx 174$ GeV, and with vectorlike quarks with SM-sized couplings constrained by the LHC to masses $\gsim700$ GeV \cite{Chatrchyan:2013uxa, ATLAS-CONF-2013-018}, requiring $y_{ij} \lsim 2$  implies an upper bound of the effective $y_b\lesssim0.5$.

\subsection{Vector and Axial-Vector Mediators}
The simple models with vector or axial-vector mediators between dark matter and the SM are already under considerable tension from collider searches and, in the case of a vector mediator, direct-detection bounds. These constraints involve only the minimal interaction; however, more complete models will typically feature couplings between the (axial-)vector mediator and other SM fields. For instance,  vector and axial-vector currents couple to both left- and right-handed fermions, and since left-handed bottom quarks are included in a weak doublet with left-handed top quarks, a coupling to tops is generically expected as well. We consider vector and axial-vector interactions that couple preferentially to third generation quarks; such couplings must align with the mass eigenstates to avoid FCNCs, and there must be additional spectator fields to cancel anomalies. We defer a discussion of such extra model-components, however, and instead focus on how the constraints in Section \ref{sec:LHC} change if the (axial-)vector mediator additionally couples to tops, since this is the most model-independent extension of the coupling to $b$-quarks in Eq.~(\ref{eq:axialvector})-(\ref{eq:vector}).

For $m_{U,V}\gtrsim350$ GeV, the collider constraints from sbottom and mono-$b$ searches are modified; the decay mode $U,V\to t\bar t$ suppresses the DM production rate. For $g_b=g_\chi$, this weakens all bounds on $\sqrt{g_\chi g_b}$ by approximately $\sqrt2$. This does not qualitatively change our conclusions, although some regions of parameter space may not be excluded until the 13 TeV running. For $g_\chi=10g_b$, however, there is no change in the bound because the mediator decays nearly always into $\chi\bar\chi$.

New production and decay modes of the mediator are now possible with $g_b=g_t$. The same mediator production processes considered in Section \ref{sec:LHC}, namely $pp\rightarrow b+U/V$, $b\bar{b}+U/V$, now lead to $t\bar t b\bar b$ production from $U/V\rightarrow t\bar t$. This modifies the total top quark production cross section. Because of the larger coupling for axial vector scenarios, requiring a contribution to the total $t\bar t$ cross section \cite{CMS-PAS-TOP-12-003} of $<10\%$ excludes axial vector masses in the range $m_U\approx 350-500$ GeV for the Fermi-favored region, while there is no bound for vector mediators from $\sigma_{t \bar t}$. 

Similarly, the mediator can now be produced via the top coupling. Production proceeds through both $pp\rightarrow t\bar{t}+Z'$, as well as by gluon fusion through a top loop. For equal couplings to top and bottom, the production rate through the $b$-coupling is much larger since there is no mass- or loop-suppression of the rate. Still, we confirmed that SM $t\bar{t}$+Higgs searches do not constrain the Fermi-favored region. We have also considered potential bounds from stop searches \cite{Chatrchyan:2013xna} and find these to be less sensitive than sbottom searches due to the smaller production cross section. 
 
%
%

\section{Conclusion}
\label{sec:conclusion}
In this paper, we have studied the direct-detection and collider constraints on 
SM singlet particles that mediate $s$-channel interactions between DM and $b$ quarks, assuming 
the mediator can decay to DM particles. For simplicity, we have emphasized only parity-conserving interactions 
 between dark and visible sectors, which restricts the class of operators whose annihilation rate is unsuppressed
 by powers of relative velocity. This is the minimal extension to the SM that suffices to explain the galactic-center gamma-ray excess identified in the Fermi-LAT data.
 Our main results are as follows:

\begin{itemize}

\item Direct-detection results from LUX disfavor a vector-vector interaction  
$(\bar \chi \gamma^\mu \chi)(\bar b \gamma_\mu b)$ that induces DM scattering off detector nuclei 
only through a $b$-quark loop; the  $m_V \gsim 300$ GeV range is ruled out.

\item Using a full collider simulation away from the contact-operator limit, we find that {\it existing} LHC sbottom
searches at $\sqrt{s} = 8$ TeV strongly disfavor the axial-vector interaction $(\bar \chi \gamma^\mu  \gamma_5 \chi)(\bar b \gamma_\mu  \gamma_5 b)$ for most combinations of perturbative couplings. While these searches have been used to constrain $t$ channel mediators that carry SM color charge (e.g. sbottoms) \cite{Berlin:2014tja},
this is the first work to highlight their sensitivity to uncolored $s$-channel mediators produced only through the interaction that also yields 
dark matter annihilation. We also find these searches to be complementary to proposed mono-$b$ + missing-energy searches \cite{Lin:2013sca},
 and present 13 TeV projections for both. 

\item The favored region for the pseudoscalar interaction $(\bar \chi \gamma^5 \chi)(\bar b \gamma_5 b)$
  is largely safe from both LHC and direct-detection bounds. Collider searches
 at 13 TeV are not sensitive to couplings for which explain the Fermi excess. 
 \end{itemize}
 
 In light of the strong constraints on the vector and axial-vector scenarios, we also considered two simple, renormalizable models 
 that give rise to pseudoscalar mediated $\chi \bar \chi \to b \bar b$ 
  annihilation. One realization involves a two-Higgs doublet model in which one couples only to 
  down-type quarks and mixes predominantly with a scalar that couples to DM.  The other 
   involves a DM coupled pseudoscalar and multiple flavors of vectorlike quarks. Integrating
   out the vectorlike states yields and effective interaction between $b$-quarks and DM and the pseudoscalar that parametrically depends on
   the ratio of Higgs VEV and vectorlike mass. Both models generically feature a suppressed  pseudoscalar-$b$ quark coupling. 


\bigskip
\section*{Acknowledgments}
We thank Wolfgang Altmannshofer, Clifford Cheung, Stefania Gori, Tracy Slatyer, and Itay Yavin for helpful conversations. EI would like to particularly thank Itay Yavin for his encouragement to publish this work. BS is supported in part by the Canadian Institute of Particle Physics. This research was supported in part by Perimeter Institute for Theoretical Physics. Research at Perimeter Institute is supported by the Government of Canada through Industry Canada and by the Province of Ontario through the Ministry of Research and Innovation.

\bibliographystyle{apsrevM}
\bibliography{HooperonDraft}

\end{document}